\documentclass{article}
\usepackage{spconf,amsmath,graphicx}
\usepackage{amsthm,amssymb}
\usepackage{mathrsfs}
\usepackage{algorithm}
\usepackage{algpseudocode} 
\usepackage{stfloats}
\usepackage{booktabs}
\usepackage{float}
\usepackage{amssymb}
\usepackage{multirow}
\usepackage{bbding}
\usepackage[table,xcdraw]{xcolor}
\usepackage[hyperfootnotes=true]{hyperref}

\title{DSCLAP: Domain-Specific Contrastive Language-Audio Pre-Training}
\name{Shengqiang Liu$^{\ast}$, Da Liu$^{\ast}$ \thanks{$^{\ast}$ Equal contribution}, Anna Wang, Zhiyu Zhang, Jie Gao, Yali Li \thanks{$^{\dagger}$ Corresponding author: yali.li@nio.com} }
\address{NIO}

\begin{document}
\maketitle
\begin{abstract}
Analyzing real-world multimodal signals is an essential and challenging task for intelligent voice assistants (IVAs). Mainstream approaches have achieved remarkable performance on various downstream tasks of IVAs with pre-trained audio models and text models. However, these models are pre-trained independently and usually on tasks different from target domains, resulting in sub-optimal modality representations for downstream tasks. Moreover, in many domains, collecting enough language-audio pairs is extremely hard, and transcribing raw audio also requires high professional skills, making it difficult or even infeasible to joint pre-training. To address these painpoints, we propose DSCLAP, a simple and effective framework that enables language-audio pre-training with only raw audio signal input. Specifically, DSCLAP converts raw audio signals into text via an ASR system and combines a contrastive learning objective and a language-audio matching objective to align the audio and ASR transcriptions. We pre-train DSCLAP on 12,107 hours of in-vehicle domain audio. Empirical results on two downstream tasks show that while conceptually simple, DSCLAP significantly outperforms the baseline models in all metrics, showing great promise for domain-specific IVAs applications.
\end{abstract}
\section{Introduction}
\label{sec:intro}

The ability to jointly understand both acoustic and semantic signals has become a  critical ability for IVAs to interpret voice signals in the physical world  \cite{Riccardi2014TowardsHP, Selfridge2015InteractTM, Huang2019ASF}. A wide range of tasks based on real-life voice have been designed to test such ability, including Multimodal Device-directed Speech Detection (MDSD)  \cite{Mallidi2018DevicedirectedUD, Norouzian2019ExploringAM, gillespie2020improving} and  Multimodal Conversational Intent Classification (MCIC) \cite{yuan2022mcic, zhang2022mintrec}. The \textit{de facto} paradigm for addressing these cross-modal tasks is to first extract acoustic features from pre-trained acoustic models  \cite{Baevski2020wav2vec2, Alec2022whisper} and textual features from pre-trained language models  \cite{Devlin2019BERTPO, Jiao2020TinyBERTDB} and then fuse them for task-specific finetune. 

Existing approaches following this paradigm have achieved sustained success due to the addition of pre-trained models, yet suffer from two main drawbacks: (1) Models are pre-trained solely on pure text or audio data,  resulting in the extracted modal representations that are disconnected in the semantic space. (2) Pre-training tasks based on generic domain data may compromise the accuracy of downstream tasks in the target domain due to the disconnections between domains. 

In this work, we propose to mitigate the above problems with a simple yet effective approach, in which we combine text and audio modalities from the target domain for domain-specific pre-training. Naïvely, we can directly refer to some popular vision-language pre-training frameworks ($\mathit{e.g.,}$ CLIP  \cite{Radford2021LearningTV} and ALIGN  \cite{Jia2021ScalingUV}) and extend them to the language-audio settings \cite{elizalde2022clap}. However, for many real-world IVA application domains, such as medical and in-vehicle, collecting and cleaning  aligned language-audio pairs is time-consuming and requires specialized annotators due to the privacy of the data, leading us always to have insufficient pairs for pre-training. On the other hand, the majority of IVAs in the real world convert the received audio signal into text through Automatic Speech Recognition (ASR) for natural language understanding. We notice that the potential of these already available language-audio pairs (probably imperfect) has not been fully explored in pre-training tasks.

To this end, we propose an approximate  approach to automatically contrast language-audio pairs for domain-specific multimodal pre-training, which we refer to as \textit{Domain-Specific Contrastive Language-Audio Pre-Training} (DSCLAP). Specifically, we train the model to align the two modalities into a common semantic space by maximizing the mutual information between the raw audio and ASR transcriptions with a contrastive learning objective and a matching objective. This \textit{ASR-transcribing}-then-\textit{training} strategy greatly reduces the need for paired data, allowing  economical end-to-end multimodal pre-training from raw audio data in the specific domain. 

We evaluated the effectiveness of our DSCLAP approach in the in-vehicle interaction domain, for which we collected over 12,107 hours of in-vehicle interaction audio data to pre-train the DSCLAP model. Experimental results show that DSCLAP significantly outperforms the baseline model and achieves the best performance not only on the MDSD task but also on the more novel yet challenging MCIC task. Notably, to our best knowledge, DSCLAP is the first model that requires only audio modality input for language-audio pre-training, which we believe is a crucial step towards domain-specific multimodal pre-training.

\section{Method}
\label{sec:method}
\begin{figure}[t]
\includegraphics[width=\linewidth]{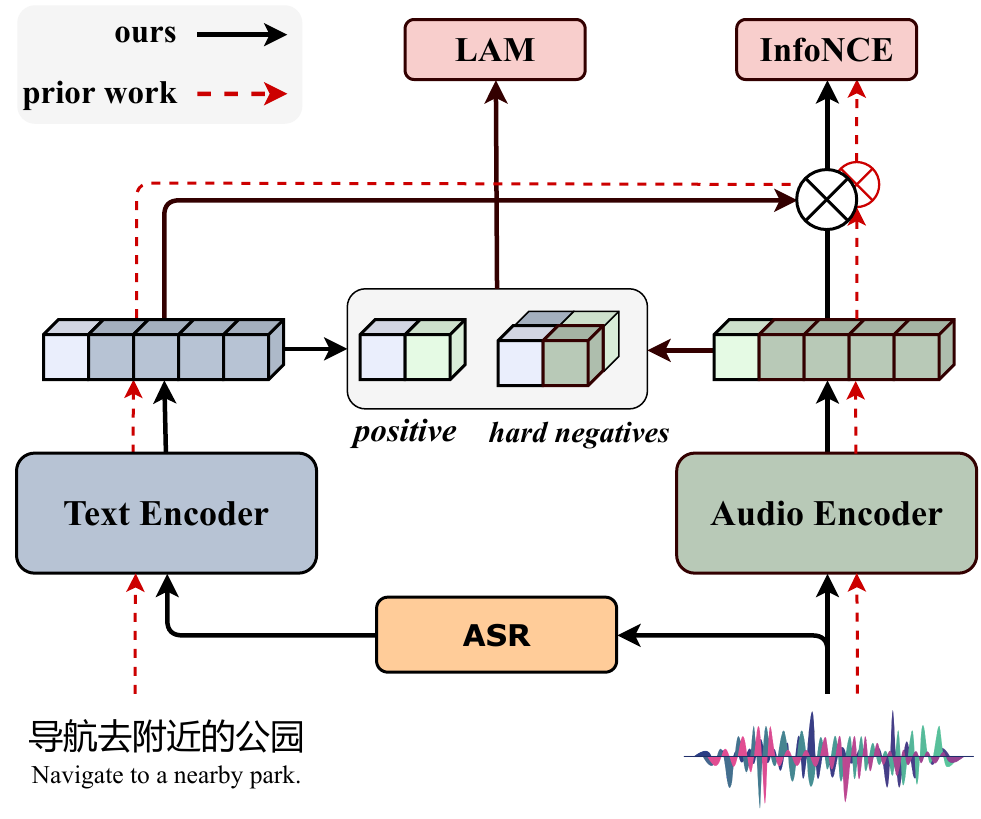}
\centering
\caption{An illustration of the DSCLAP framework. In contrast to prior work  \cite{elizalde2022clap} that leverages pre-prepared language-audio pairs for contrastive learning pretraining (\textit{red dashed arrows}), our DSCLAP (\textit{black hard arrows}) requires only raw audio inputs. Besides the standard InfoNCE loss, inspired by \cite{Zeng2022CrossViewLM}, we introduce a \textbf{L}anguage-\textbf{A}udio \textbf{M}atching (LAM) objective to achieve more effective contrastive learning.}
\label{fig1}
\end{figure}

We propose a new framework that enables multimodal pre-training with only raw audio modality input. The overall architecture of the framework is shown in Fig. \ref{fig1}.  Specifically, we first translate audio signals to text with an ASR system. Then, we encode these two modalities and fuse their representations with a cross-modal modeling layer. Finally, we train the model with a contrastive learning objective and a language-audio matching objective.

\subsection{Embeddings}

For encoding audio frames we utilize a pre-trained Wav2vec2 whose architecture follows the work in  \cite{Baevski2020wav2vec2}. Let the raw waveform be $\textbf{\footnotesize{X}}_{a}$ and $\mathcal{F}_{a}$ represent the Wav2vec2 encoder.  For $N$ waveforms in a batch, the final acoustic representations are calculated as follows:

\begin{equation}\label{eq1}
h^{(i)}_{a} = \text{Pooling}(\mathcal{F}_{a}(\textbf{\footnotesize{X}}_{a}^{(i)})), ~~ i\in\{1,2,...,N\}
\end{equation}
where $h^{(i)}_{a}\in\mathbb{R}^{d}$  is the $i$-th audio representation of dimensionality ${d}$. A $Pooling$ layer ($\mathit{i.e.,}$ mean-pooling) is applied to aggregate the frame-level features into an utterance-level representation.

We choose TinyBERT  \cite{Jiao2020TinyBERTDB} as the text backbone network, a Transformer model trained with attention-based distillation and hidden states-based distillation. Let the text be $\textbf{\footnotesize{X}}_{t}$ and $\mathcal{F}_{t}$ represent the TinyBERT encoder. Similar to the audio encoding process, we can obtain the linguistic representation $h^{(i)}_{t}$:

\begin{equation}\label{eq1}
h^{(i)}_{t} = \text{Pooling}(\mathcal{F}_{t}(\textbf{\footnotesize{X}}_{t}^{(i)})), ~~ i\in\{1,2,...,N\}
\end{equation}

\subsection{Contrastive Learning}

Given a representation pair ($h^{(i)}_{a}$ , $h^{(i)}_{t}$), following CLIP  \cite{Radford2021LearningTV}, we use InfoNCE loss to  measure dependencies between the audio and text modality. Specifically, we project $h^{(i)}_{a}$ and $h^{(i)}_{t}$ onto the latent embedding space. Then we maximize the similarity between the pair of audio and text representation ($\mathit{i.e.,}$ positive pair) while minimizing the similarity between the negative pairs as follows:

\begin{gather}\label{eq2}
\textbf{z}^{(i)}_{a} = \phi_{a}(h^{(i)}_{a}),  ~~\textbf{z}^{(i)}_{t} = \phi_{t}(h^{(i)}_{t}) \\
\mathcal{L}_{a}= -  \log  \frac{\mathrm{exp}(\mathrm{sim}(\textbf{z}^{(i)}_{a},\textbf{z}^{(i)}_{t})/\tau)}{\sum_{\textbf{z}^{(j)}_{t}\in\mathcal{B}}\mathrm{exp}(\mathrm{sim}(\textbf{z}^{(i)}_{a},\textbf{z}^{(j)}_{t})/\tau)} \\
\mathcal{L}_{t}=- \log  \frac{\mathrm{exp}(\mathrm{sim}(\textbf{z}^{(i)}_{t},\textbf{z}^{(i)}_{a})/\tau)}{\sum_{\textbf{z}^{(j)}_{a}\in\mathcal{B}}\mathrm{exp}(\mathrm{sim}(\textbf{z}^{(i)}_{t},\textbf{z}^{(j)}_{a})/\tau)}
\end{gather}
 where $\phi_{a}$ and $\phi_{t}$ are learned fully-connected projection functions, $\textbf{z}^{(i)}_{a}$ and $\textbf{z}^{(i)}_{t}$ are hidden representations of the $d$ dimension, $sim()$ is a similarity function ($\mathit{e.g.,}$ dot product). Furthermore, $\mathcal{B}=\{\textbf{z}^{(1)}_{\beta},\textbf{z}^{(2)}_{\beta},...,\textbf{z}^{(N)}_{\beta}\}$ ($\beta\in\{a,t\}$) is a set of hidden representations, which contains a positive sample $\textbf{z}^{(i)}_{\beta}$ and $\text{N-1}$ negative samples ( $\mathcal{B}-\textbf{z}^{(i)}_{\beta}$).

\subsection{Language-Audio Matching}
Contrasting learning objective often requires a large batch size, which in turn increases a tremendous computational resource cost. To cut down the heavy resource dependency, we introduce a language-audio matching objective with hard negative sampling \cite{Zeng2022CrossViewLM} to allow us to conduct efficient multimodal representation alignment with limited resources. Specifically,  for a given fixed multimodal representation pairs ($\textbf{z}^{(i)}_{a}$ , $\textbf{z}^{(i)}_{t}$), a set of hard negative examples \{ ($\textbf{z}^{(i)}_{a}$ , $\textbf{z}^{(k)}_{t}$),  ($\textbf{z}^{(k)}_{a}$ , $\textbf{z}^{(i)}_{t}$)\}, are sampled from the top-K most similar representations in the training batch. In other words, we use carefully selected hard negative samples as an additional negative sample for contrastive learning as follows:

\begin{gather}\label{eq2}
\mathcal{L}_{a^{*}}= -  \log  \frac{\mathrm{exp}(\mathrm{sim}(\textbf{z}^{(i)}_{a},\textbf{z}^{(i)}_{t})/\tau)}{\sum_{ \textbf{z}^{(\bar{j})}_{t} \in \{ \textbf{z}^{(i)}_{t}, ~ \textbf{z}^{(k)}_{t}\}} \mathrm{exp}(\mathrm{sim}(\textbf{z}^{(i)}_{a},\textbf{z}^{(\bar{j})}_{t})/\tau)} \\
\mathcal{L}_{t^{*}}=- \log  \frac{\mathrm{exp}(\mathrm{sim}(\textbf{z}^{(i)}_{t},\textbf{z}^{(i)}_{a})/\tau)}{\sum_{ \textbf{z}^{(\bar{j})}_{a} \in \{ \textbf{z}^{(i)}_{a}, ~ \textbf{z}^{(k)}_{a}\}  }\mathrm{exp}(\mathrm{sim}(\textbf{z}^{(i)}_{t},\textbf{z}^{(\bar{j})}_{a})/\tau)}
\end{gather}

\subsection{DSCLAP objective}
Incorporating the loss on the contrastive learning and the cross-modal matching introduced above, we estimate the parameters of the pre-train model $\theta$ by minimizing the following objective:

\begin{equation}
    \min_{\theta} \lambda (\mathcal{L}_{a}+\mathcal{L}_{t}) + \gamma(\mathcal{L}_{a^{*}}+ \mathcal{L}_{t^{*}})
\end{equation}
where $\lambda$ and $\gamma$ are hyperparameters that determine the contribution of each regularization component. For all the experiments, we set $\lambda$ and $\gamma$ as $0.5$, which we searched through cross-validation.

\section{Experiments}
\label{sec:experiments}

\subsection{Datasets}

\subsubsection{Pre-training}

We pre-train our DSCLAP in the in-vehicle domain. Specifically, we collect a total of 12,107 hours of training corpus, which contains more than 20M unlabeled raw audio from more than 250K vehicles via an in-vehicle IVA. We convert these audios to text using the 
\textit{iFLYTEK cloud speech service}\footnote{www.xfyun.cn/services/lfasr}, which provides the current state-of-the-art ASR system in China, achieving an approximate 18.7\% character error rate in the training corpus. Note that although the language-audio pairs used for pre-training may not be perfect, DSCLAP can still learn appropriate representations from positive and negative examples through comparative learning. Moreover, fine-tuning on downstream tasks can further correct the effects of ASR errors. We will discuss this in Section 4.


\subsubsection{Downstream Tasks}
We evaluate DSCLAP on two downstream tasks, corresponding to the two hot research fields: 1) \textbf{Multimodal Device-directed Speech Detection} (MDSD), which allows IVAs to continuously listen to the user within a predefined period without repeatedly detecting a wake-up-word or trigger phrase in human-computer interaction \cite{gillespie2020improving, vilaysouk2021improving}. 2) \textbf{Multimodal Conversational Intent Classification} (MCIC), which aims to jointly understand user intent through information from both text and audio modalities  \cite{yuan2022mcic, zhang2022mintrec}. Both MDSD and MCIC are classification tasks, where MDSD includes two categories of device-directed or non-device-directed, and MCIC includes 15 intents, such as playing music, navigating, calling, and chatting. The details of training data, validation data, and test data splitting are shown in Table \ref{table1}.

\begin{table}[htbp]
\centering
\resizebox{\linewidth}{!}{
\begin{tabular}{cc|ccc|c}
\hline
\toprule
\multicolumn{2}{c|}{Task}& \multicolumn{1}{c|}{Train} & \multicolumn{1}{c|}{Valid} & Test  & Count  \\ \hline
\multicolumn{1}{c|}{\multirow{2}{*}{MDSD}} & \textbf{ASR-only}    & 10,000   & 5,000    & 4,800 & 19,000 \\ \cline{2-6} 
\multicolumn{1}{c|}{}    & \textbf{ASR\&Manual} & 2356     & 337      & 674   & 3,367  \\ \hline
\multicolumn{2}{c|}{MCIC} & 6,400        & 2,000        & 1,800     & 10,200      \\ \bottomrule
\end{tabular}
}
\caption{Statistical information and data partition of datasets used in this paper.}
\label{table1}
\end{table}

\subsection{Implementation Details}
We use AdamW  optimizer for pre-training and the learning rate is initialized to {2e-5}. The the temperature $\tau$ of contrastive loss is initialized to $log(1\//0.07)$. The maximum sequence length is set to {16} and {80,000} for texts and audio waveforms respectively. Our model is implemented in PyTorch Lightning  and transformers libraries. We train it for {20} epochs with mixed precision, on {8} NVIDIA RTX 3090 GPUs with a batch size of {64} per GPU. The whole training process takes {7.5} days to complete. Notably, for a fair comparison, we used five random seeds (i.e. 1, 12, 123, 1234, and 12345) to trigger experiments in both of the downstream tasks and report the average performance of the five experiments for each task.

\section{RESULTS AND ANALYSIS}
\label{sec:typestyle}

\subsection{Quantitative Results}

We combined several recent state-of-the-art unimodal pre-train models to benchmark their performance on domain-specific downstream tasks and report the results of the two best-performing models: (Whisper \cite{Alec2022whisper}, BERT \cite{Jiao2020TinyBERTDB}) and (Wav2vec2 \cite{Baevski2020wav2vec2}, BERT \cite{Jiao2020TinyBERTDB}), where the better-performing (Wav2vec2, BERT) is set as a baseline backbone network (annotated by \textbf{Base}). Additionally, we also introduce two novel competitive optimization methods (Rdrop \cite{wu2021r} and OGM-GE \cite{peng2022balanced}) on the \textbf{Base} model.

The comparative results are presented in Table \ref{table2}. In both datasets, DSCLAP achieves the best performance and surpasses the variants across all metrics (MDSD and MCIC combined). This confirms that our method can be better suited to domain-specific multimodal tasks. We observe that the proposed model significantly reduced the \textit{false rejection rate} (FRR) in the MDSD task from 5.35\% to 2.77\%, which is of great benefit to IVAs because it will enable IVA to respond to more user requests without wake-up steps.

\begin{table}[t]
\centering
\resizebox{\linewidth}{!}{
\begin{tabular}{l|cc|cc}
\hline
\toprule
\multirow{2}{*}{Model} & \multicolumn{2}{c|}{MDSD}     & \multicolumn{2}{c}{MCIC}         \\ \cline{2-5} 
 & ACC   & FRR & ACC   & F1             \\ \hline
(Whisper \cite{Alec2022whisper}, BERT \cite{Jiao2020TinyBERTDB})  & 91.25 & 10.82 & 87.57 & 86.98          \\
(Wav2vec2 \cite{Baevski2020wav2vec2}, BERT \cite{Jiao2020TinyBERTDB})$_\textbf{Base}$   & 92.48 & 6.11 & 87.93 & 87.52          \\
Base-OGM-GE \cite{peng2022balanced}   & 92.51 & 5.93 & 88.90 & 88.47          \\
Base-RDrop \cite{wu2021r}    & 93.57 & 5.35 & 88.54 & 88.01          \\ \hline
DSCLAP   & \textbf{95.06} & \textbf{2.77} & \textbf{89.98} & \textbf{89.76} \\
\bottomrule
\end{tabular}
}
\caption{Experimental results on MDSD \textbf{ASR-only} dataset and MCIC dataset. The best results are highlighted in bold.}
\label{table2}
\end{table}

\subsection{Ablation Study}
\begin{figure}[t]
\includegraphics[width=0.85\linewidth]{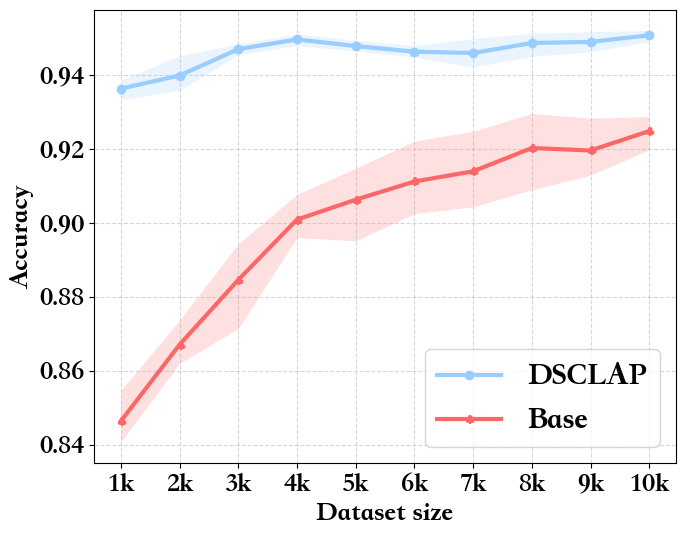}
\centering
\caption{ Accuracy curves for MDSD on \textbf{ASR-only} dataset w.r.t. different size of training data. Values are for five runs across random seeds.}
\label{fig2}
\end{figure}

\noindent
\textbf{4.2.1. Do we need a larger training data size for finetuning?}
We compare the impact of different training data sizes $M\in\{1\text{\footnotesize{K}},2\text{\footnotesize{K}},...,10\text{\footnotesize{K}}\}$ on the Base model and our proposed model on the MDSD task, results shown in Fig. \ref{fig2}. We observe that the accuracy curves of both models showed an increasing trend with the increase in training data. However, with the decrease in data size, the accuracy of the Base model declined dramatically. In contrast, our model is still stable at an accuracy higher than 93.33\% with $1\text{\footnotesize{K}}$ training data, suggesting that DSCLAP can help the Base model better adapt to few-shot data scenarios and provide a possible solution for cold start problems with the domain-specific model. \vskip 0.2cm

\noindent
\textbf{4.2.2. Is pre-training with imperfect language-audio pairs feasible? } 
We also experiment on different text inputs under the same audio source, with results shown in Table \ref{table3}. We observe that our DSCLAP outperforms the Base model by 2.26\% on manually transcribed data. This demonstrates that although there is some noise (ASR transcription error) in the pre-training data, DSCLAP can effectively learn the alignment information between the two modalities by contrastive learning.  the model improves from 2.26\% to 5.37\% when comparing the ASR transcript pairs to manual transcript pairs fine-tuning results. This is an encouraging result that enables us to better apply it to IVAs since the text data of IVAs are always transformed by real-time ASR systems. Additionally, we found that DSCLAP did not perform as well as ASR transcriptions on manual transcriptions. This may be caused by the inconsistency between the downstream task and the pre-training task input. Because DSCLAP is trained on a large amount of ASR transcription pairs, it may be more suitable for downstream tasks that use ASR transcription pairs as input. \vskip 0.2cm

\begin{table}[t]
\centering
\begin{tabular}{c|c|c|c}
\hline
\toprule
Source & Base  &  DSCLAP  &  $\bigtriangledown$   \\ \hline
ASR    & 89.82 & 95.19 & {\color[HTML]{009901} $\uparrow$ 5.37} \\ \hline
Manual & 91.98 & 94.24 & {\color[HTML]{9B9B9B}  $\uparrow$ 2.26} \\    \bottomrule      
\end{tabular}
\caption{Comparison of performance of models in ASR transcript pairs and manual transcript pairs on \textbf{ASR \& Manual} dataset.  The col with $\bigtriangledown$ means the improvements of our model compared to the Base model in accuracy.}
\label{table3}
\end{table}

\noindent
\textbf{4.2.3. Base encoders \textit{vs.} DSCLAP encoders.}
To investigate the difference in generalization between the Base and DSCLAP on encoders, we compare the two models with variants that freeze portions of the parameters. Results are shown in Table \ref{table4}. Overall, our DSCLAP achieves better performance than the base model on all variants, showing the benefits of domain-specific pre-training. For DSCLAP, we observe that training only with the text encoder parameters results in the worst performance, contrary to previous works \cite{elizalde2022clap}. This is possibly because the text used for fine-tuning is noisy and imperfect ASR transcriptions, instead of manually transcripts in previous work.

\begin{table}[t]
\centering
\begin{tabular}{c|c|c|c|c}
\hline
\toprule
$\mathcal{F}_{a}$ & $\mathcal{F}_{t}$ & Base  &  DSCLAP & $\bigtriangledown$ \\ \hline
{\footnotesize \XSolidBrush}  & {\footnotesize \XSolidBrush}  & 85.08 & 93.66   & {\color[HTML]{009901}  $\uparrow$ 8.58}    \\ 
{\footnotesize \XSolidBrush}  & {\footnotesize \CheckmarkBold }  & 88.63 & 90.67   &  {\color[HTML]{9B9B9B}  $\uparrow$ 1.94}  \\ 
{\footnotesize \CheckmarkBold }  & {\footnotesize \XSolidBrush}  &  91.17 & 94.95   &  {\color[HTML]{9B9B9B}  $\uparrow$ 3.78} \\ \hline
{\footnotesize \CheckmarkBold }  & {\footnotesize \CheckmarkBold }  & \textbf{92.48} & \textbf{95.06}   & {\color[HTML]{9B9B9B}  $\uparrow$ 2.58}         \\ \bottomrule   
\end{tabular}
\caption{\text{Base encoders} \textbf{vs.} \text{DSCLAP encoders.} {\footnotesize \CheckmarkBold } indicates the parameters of the corresponding encoder are trainable.}
\label{table4}
\end{table}

\section{CONCLUSION}
\label{sec:page}
In this work, we aim to address the domain-specific multimodal pre-train problem that lacks paired training data. To this end, we propose DSCLAP, which uses an ASR system to automatically generate text from raw speech signals and builds a multimodal pre-train model with this language-audio pair. Compared with previous works, DSCLAP greatly reduces the cost of data collection and cleaning, allowing economical end-to-end learning from raw audio and ASR transcriptions. By pre-training on 12,107 hours of in-vehicle domain audio data, DSCLAP provides more efficient domain-specific modal representations, helping the model achieve state-of-the-art results on downstream tasks. 

DSCLAP demonstrates that efficient multimodal pre-training can be accomplished using only raw audio signals. We believe that the performance of DSCLAP on downstream tasks could lead to many exciting possibilities for its future applications in IVA. We hope that the training strategy of DSCLAP could encourage researchers for further explorations on different domains.

\vfill\pagebreak


\bibliographystyle{IEEEbib}
\bibliography{main}

\end{document}